# Task Allocation in Mobile Crowd Sensing: State of the Art and Future Opportunities

Jiangtao Wang, Leye Wang, Yasha Wang, Daqing Zhang, and Linghe Kong

*Abstract*—Mobile Crowd Sensing (MCS) is the special case of crowdsourcing, which leverages the smartphones with various embedded sensors and user's mobility to sense diverse phenomenon in a city. Task allocation is a fundamental research issue in MCS, which is crucial for the efficiency and effectiveness of MCS applications. In this article, we specifically focus on the task allocation in MCS systems. We first present the unique features of MCS allocation compared to generic crowdsourcing, and then provide a comprehensive review for diversifying problem formulation and allocation algorithms together with future research opportunities.

## I. INTRODUCTION

Urban sensing is crucial for understanding the current status of a city in many aspects (e.g., air quality, traffic status, noise level, etc.). With the development of Internet of Things (IoT), mobile internet and cloud computing, we now have various ways to collect urban information [1, 2, 3]. Among them, the prevalence of mobile devices and the increasing smart sensing requirements in the city have led to an alternative or complementary approach for urban sensing, called Mobile Crowd Sensing (MCS) [4]. Similar concepts include participatory sensing [5], location-based/mobile/spatial crowdsourcing [6], collaborative sensing [7], and so forth.

MCS leverages the inherent mobility of mobile users (i.e., participants or workers), the sensors embedded in mobile phones and the existing communication infrastructure (Wi-Fi, 4G/5G networks) to collect and transfer urban sensing data. MCS has enabled diverse applications, such as air quality monitoring [34], noise level sensing [19], queue time estimation [68], risky mountain trail detection [78], and so forth. Compared to wireless sensor networks (WSN), which are based on specialized sensing infrastructures, MCS is less costly and can obtain a higher spatial-temporal coverage. As a result, the emergence of MCS has expanded the scope of IOT, where the "things" are not only limited to physical objects (i.e., they also include human and their carried mobile devices).

The connection between tasks and workers is crucial for the success of MCS applications. The simplest way is that the organizers publish various MCS tasks and workers select tasks themselves based on their location and preferences (e.g., Medusa [9] and PRISM [10]), which is called the pull mode. The pull mode is easy to implement. However, for the pull mode, the cloud server does not have any control over the tasks assignments. Since workers select tasks based on their own preference or goals (e.g., nearby, easy, or high payment), the overall performance may not be globally optimized. For example, some sensing tasks have few participants so that the sensing quality is low, while others may have too many which leads to redundant sensing data.

Therefore, it is a promising technical alternative that the server automatically assigns sensing tasks to workers according to the system optimization goals (e.g., maximizing the sensing quality while ensuring the budget constraints), which is called the push mode. In recent years, the studies for automatic MCS task allocation becomes a hotspot in research communities such as ubiquitous computing, social computing, cooperative computing, and computer network.

There are some tutorials or surveys (e.g., [1] and [25]) for MCS in recent years. The scope of these papers is for the entire research community of MCS, which discuss different aspects and research issues in this field to give us an overview picture of MCS. However, as these survey papers mainly focus on the general and overall research picture and roadmap of MCS, none of them summarize and discuss the research problem of sensing task allocation in details and systematically. Especially as task allocation is one of the hottest research topics in MCS where there are still continuous achievements published in top venues across various areas in recent years (e.g., ICDE [73], UbiComp [33], CSCW [52], WWW[66] , IEEE TMC [74], IEEE TIST [75]), a tutorial or survey devoted to summarizing its up-to-date research results is even desirable. To this end, in this article, we specifically focus on the task allocation problem in MCS and provide a comprehensive review with future research opportunity.

The possible inspiration derived from this article consists of the following aspects:
1) We analyze the unique factors or features in MCS in addition to general crowdsourcing, which can reveal why the traditional task assignment methods for

Jiangtao Wang, Yasha Wang, and Daqing Zhang are with computer science department in Peking University, Beijing, China, 100871 (e-mail: {jiangtaowang, wangyasha}@ pku.edu.cn, dqzhang@sei.pku.edu.cn).

Leye Wang is with computer science and engineering department, Hong Kong University of Science and Technology, Hong Kong SAR, China (e-mail: wly@cse.ust.hk).

Linghe Kong is with Department of Computer Science and Engineering at Shanghai Jiao Tong University, Shanghai, China.

crowdsourcing cannot be directly utilized to tackle the task allocation problem in MCS.
2) We present and summarize different types of problem formulation in MCS task allocation and corresponding algorithms, which help the researchers or engineers to quickly identify the subset of studies and provide guidance or inspiration when designing and implementing the MCS systems or applications.
3) We discuss some potential research directions and proposals, which aims to consider more practical issues in MCS task allocation.

## II. PRELIMINARY FOR MOBILE CROWD SENSING

### A. Crowdsourcing and Mobile Crowd Sensing

The term "crowdsourcing" was coined by Jeff Howe and Mark Robinson in [11] to describe how businesses were using the Internet to outsource work to the crowd. The basic idea of crowdsourcing is to leverage the power of crowd to collaboratively complete a complex task, where each individual (called "worker") only completes much easier micro-tasks. In recent years, crowdsourcing-based systems are widely used in many domains [12, 13, 14, 15, 16] (see Fig. 1), and the intersection between crowdsourcing and these tradition research areas gives rise to a new research topic.

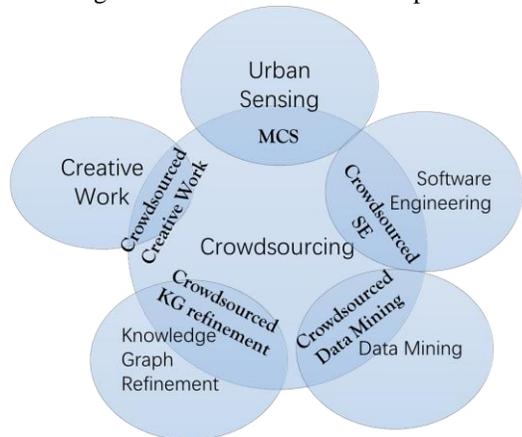

Fig. 1 Crowdsourcing used in different domains

The popularity of mobile devices and the increasing sensing requirement in the city enable a subclass of crowdsourcing called the mobile crowd sensing (MCS) [1]. Similar to the notion of participatory sensing [17] and human-centric computing [18], MCS refers to the sensing paradigm in which users with sensor-rich mobile devices collect and contribute data in order to enable various applications. In an MCS system, there are two key players, i.e., workers (or participants) who collect and report sensing data through mobile device, task organizers who manage and coordinate whole MCS process. Various kinds of MCS applications have been proposed and implemented in both academic and industry areas, such as environmental applications [19,20], infrastructure applications [21,22,8], and social applications [23,24]. For detailed introduction and classification about various MCS applications, interested readers can refer to a recent survey paper [30].

### B. Life-Cycle of MCS and Research Issues

The life-cycle of MCS can be divided into four stages: task creation, task allocation, task execution and data aggregation. The main functionality and research issues of each stage are briefly described as follows: (1) *Task Creation*: The MCS organizer creates an MCS task through providing the workers with the corresponding mobile phone applications. In this stage, the key research issue is how to improve the efficiency of MCS task creation, especially for those who do not have professional programming skills [9][26]. (2) *Task Allocation*: After the organizer creates an MCS task, the next stage is task allocation, in which the application or a public platform recruits workers and assigns them with sensing tasks. The key research issue at this stage is how to optimize the task allocation with the consideration of diverse factors, such as spatial coverage, incentive cost, energy consumption, and task completion time [47,52]. (3) *Task Execution*: Once receiving the assigned micro sensing tasks, the workers complete them within a pre-defined spatial-temporal scale (i.e., time duration and target region). This state includes sensing, computing, and data uploading. How to save energy consumption is the major research issue in this stage [28,29]. (4) *Crowd Data Integration*: This stage aggregates the reported data from the crowd according to the requirement of task organizers. The key issue in this stage is how to infer missing data and provide a complete spatial-temporal picture of the target phenomenon (e.g., the real-time air quality map in the city) [40, 41].

In this article, we specifically focus on the task allocation stage. We first present the unique features of MCS allocation compared to generic crowdsourcing. Then, we provide a comprehensive review for diversifying problem formulation and allocation algorithms together with future research opportunities.

## III. SPECIFIC FACTORS IN MCS TASK ALLOCATION

### A. Overview

MCS is the special case where the idea of "crowdsourcing" is used in urban sensing scenarios. Task allocation of MCS shares some common concerns or factors with general crowdsourcing tasks (e.g., article writing or image classification) [31]. For example, both general crowdsourcing and MCS consider incentive models and budget constraints in task allocation strategies. On the other hand, MCS has its own unique features which differ from general crowdsourcing. To this end, we provide the comparative schemas of general crowdsourcing and MCS in Fig. 2, where the green color labels the unique factors of MCS.

Essentially, the unique characteristic of MCS lies in the aspects of *mobility* and *sensing*. Thus, we elaborate the MCS-specific factors or features from these two aspects.

**Mobility-Relevant Features.** Different from general crowdsourcing tasks, MCS requires the workers to complete sensing tasks in certain locations, because the sensing results are location-dependent (e.g., air quality, noise level, and traffic congestion status). This characteristic leads to the "participatory mode" and "location privacy" features in Fig. 2.

First, based on how the workers move to the locations for sensing, we can divide MCS task allocation into two participation modes (i.e., participatory or opportunistic). Second, since MCS usually targets at collecting *spatial* data all across a city, location privacy should be carefully preserved. In addition, spatial-temporal models usually need to be considered in the sensing quality metric of MCS, but rarely in the task quality of general crowdsourcing.

**Sensing-Relevant features.** Different from general crowdsourcing, MCS always targets at urban sensing tasks. First, the execution of sensors and localization modules introduces much more energy consumption into MCS than general crowdsourcing. The energy consumption has a direct impact on the battery life of a worker's smartphone. If the energy consumption of an MCS task is too high, it will severely reduce the mobile phone users' willingness of becoming a crowd worker. Therefore, it is important to control the energy consumption of workers in the MCS systems, which is also labeled as a unique feature in Fig. 2. Second, many MCS tasks need to invoke phone-embedded sensors for task completion, but the set of sensors for each worker may be different as they hold various brands and models of smart devices. Thus, the "sensor type requirement" should be particularly considered in the task allocation of MCS.

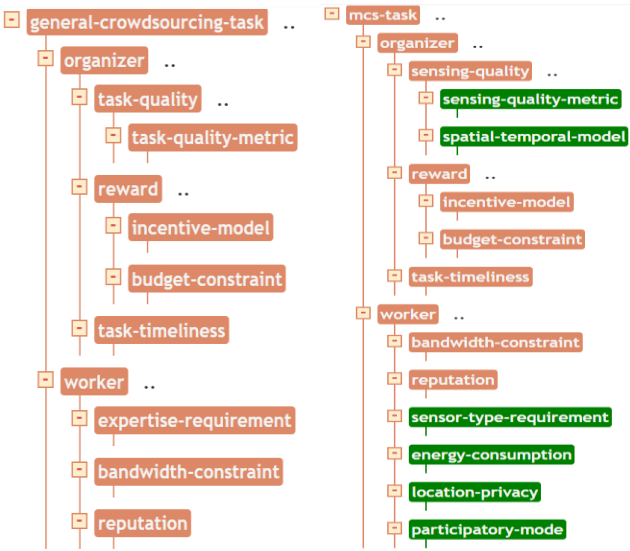

Fig. 2 Comparative schemas of general crowdsourcing and MCS (left: general crowdsourcing, and right: MCS. Green one labels the unique factors for MCS compared to general crowdsourcing).

As the worker and organizer are the key roles in MCS, we divide the above MCS-specific factors into two categories from the perspective of worker and organizer, respectively.

### B. Worker-Side Factors.

**Workers' Participation Mode.** (1) *Participatory Mode*. This mode requires the workers to change their original routes and specifically move to certain places to complete MCS tasks [32, 33], and its advantage is that it can guarantee task completion. However, since workers need to deviate from their original routines and travel to task locations, it incurs extra travel cost and can be intrusive to the workers. It also increases the task organizers' incentive cost, since the task organizers usually have to pay extra incentive rewards to compensate for the traveling cost of the workers. (2) *Opportunistic Mode*. For this mode, workers can complete tasks unintentionally during their daily routines without the need to change their routes [34, 35]. The opportunistic mode does not require knowledge of the workers' intended travel routes, so it is less intrusive for the workers and less costly for the task organizers. However, the sensing quality of the assigned tasks depend heavily on the workers' routine trajectories. For tasks that are located at places visited by few or even no workers, their sensing quality can be very poor.

**Location Privacy**. There have been proposed a spectrum of location privacy preserving techniques for location-based services, and many of them have also been successfully adopted in MCS [67]. Different privacy mechanisms may have their own metrics to quantify the privacy protection effect. For example, the *cloaking* mechanism is often designed based on the *k-anonymity* metric, i.e., ensuring that a user's reported location is same as the other k-1 users (i.e., a user is indistinguishable from the other k-1 users) [70]; ε-differential-privacy is a location obfuscation scheme to protect users' real locations, which is able to bound the adversary's posterior knowledge improvement over his prior knowledge about a user's location, while ε can be set by users' privacy preferences [66]. In other words, if an adversary foreknows that a user has a probability of P in a location L, with the ε-differential-privacy protection, the adversary's confidence probability of the user at L will not be larger than C*P after observing the user's obfuscated location, where C is a constant determined by ε. As location privacy protection mechanisms generally include noises added into participants' locations, it will bring novel challenges for task allocation, e.g. locations of users' uploaded data become somehow uncertain [71] and the distance between users and task locations cannot be precisely measured [66]. Then, finding the optimal privacy mechanism, where the loss of task allocation efficiency is minimized, becomes rather important.

**Energy Consumption.** Several methods proposed in for mobile phone sensing [36] can be directly used to reduce the energy consumption for an individual worker, which are mainly adopted in the sensing and data uploading phase of MCS. Additionally, we can further optimize the overall energy consumption by designing more sophisticated task allocation mechanism [27]. In this article, we focus on how to take the energy consumption concern into consideration in the task allocation phase.

### C. Organizer-Side Factors

**Spatial-Temporal Model.** Different from general crowdsourcing, task organizers in MCS can obtain a spatial-temporal overview of the environment in the target area (e.g., air quality map in Beijing) by collecting sensor readings from mobile users. The most common way of modeling the time and space in MCS is to divide the entire sensing areas and time period into some equal-size subareas (1km*1km) and equal-length time slots (1hour per slot), so that we can get a number of spatial-temporal cells [34,35,51,52]. Another way is regarding the sensing target as a POI (Point-of-Interest) with a given range (e.g., a circle with 100m radius). If a mobile user moves inside such a range of a POI, he/she can collect the sensing data at this point [55,56,57,58]. Most of the general crowdsourcing tasks do not consider the location of the workers and sensing cycles. But for some, such as Internet quality measurement [59], the time and location of the reported network quality information is also considered. However, their spatial-temporal models are quite different, where the topology (the spatial model) and peak hours (the temporal model) of the network are considered to measure the service quality.

**Sensing Quality**. For MCS task allocation, the quality of sensing data is a primary concern for the task organizer. Thus, how to model or quantify the quality of sensing task in MCS should be considered. The sensing quality metric can be divided into the following two types. (1) Spatial-temporal coverage based metrics. One naive to measure the quality of sensing data is based on the number of collected data samples. Accordingly, a common metric to measure the sensing quality of an MCS tasks is the spatial-temporal coverage, i.e., how many subareas can be covered by the sensing data collected [34,35,51,52]. It is also different in defining whether a subarea is "covered" or not. To simplify the problem, earlier research works always assume that if one subarea gets one data sample, it is regarded as "covered" in this time slot. However, recent studies such as [52] assume that at least a number of samples is needed (i.e., the minimum threshold) to guarantee the reliability of collected data. Then, if the minimum requirement is met, the coverage quality would increase as the number of samples increases until reaching to a certain degree (i.e., the maximum threshold). (2) Sensor data value based metrics. Due to the temporal and spatial correlations in the MCS systems, the sensor readings of some spatial-temporal cells can be inferred from the others. In this case, another typical way to quantify the sensing quality is to infer the data of sub-areas without sensor readings and then compute the inference error [40, 41, 42]. Especially, the average inference error among all the sub-areas is also often used as a quality metric [40, 41] for continuous sensing values (e.g., temperature), while average classification error is used for classification-based sensing values (e.g., air quality level) [42].

## IV. PROBLEM FORMULATION IN MCS TASK ALLOCATION

Task allocation for MCS tasks is commonly formulated as the mathematical optimization problems with various goals and constraints. We classify and summarize the state-of-the-art research works from the following perspectives (see Fig. 3).

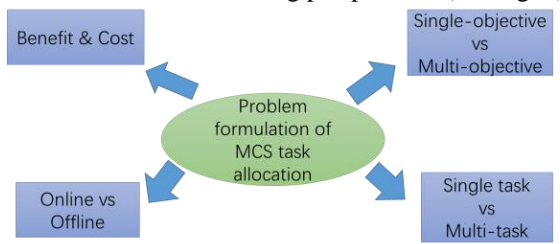

Fig. 3 Classification of problem formulation in MCS task allocation

### A. Benefit and Cost

The essence of MCS task allocation is to achieve the tradeoff between several opposing factors, which is divided into two classes called *benefit* and *cost*. The benefit is defined as the sensing quality of an MCS task, which may be measured by different metrics (i.e., spatial-temporal coverage based metrics and sensor data value based metrics described in Section III.C). However, in order to achieve higher benefit (or sensing quality), some overhead factors, which is called *cost* is this article, should be taken into account. The cost factors mainly include incentive cost, risk of location privacy leak, energy consumption, and so forth. We classify the existing work as follows, which is based on the type of cost in the formulated optimization problems.

First, to improve the sensing quality of MCS tasks, the naïve way is to assign tasks to as many workers as possible. However, too many task assignments will lead to the increase in incentive cost. Thus, sensing quality and cost are two opposing factors, and managing the trade-off between them through task allocation is a fundamental and crucial research problem. Several research studies were proposed recently which aimed at either maximizing the sensing quality with budget constraints (e.g., [43,44,45]), or minimizing the incentive cost while guaranteeing a minimum level of sensing quality (e.g., [34,46,47]).

Second, sensing quality and location privacy are often two conflicting-objectives in task allocation optimization. To protect mobile users' locations, their actual locations are often perturbed or obfuscated before being uploaded to the server. Usually, the higher protection effect is desired (i.e., the location is more inaccurate), the lower sensing quality could be obtained. One of the most commonly used location privacy protection methods belongs to the category of cloaking, where a user's fine-grained location is down-graded to a coarse-level region [67]. More recently, differential privacy is applied in MCS to provide a theoretical privacy guarantee regardless of any adversary's prior knowledge about his victim user's location distribution [66]. To obtain the highest sensing quality while ensuring privacy protection effect, many researchers have formulated an optimization problem for task allocation, where privacy protection effect is often regarded as constraints and sensing quality as optimization objectives [66, 68].

Third, several methods can be utilized to reduce the energy consumption in MCS. For example, in sensing data collection phase, the authors in [36] design new methods using a set of energy-efficient sensors to replace the traditional approaches consisting of more energy consuming sensors, or dynamically adjust the data collection frequency to do tasks more efficiently. In the data transferring phase, low-power wireless communication network (e.g., Wi-Fi) is utilized to upload data, rather through 3G/4G [48], or upload data to the server when users established the Internet connections for other applications, called piggyback [49]. However, all the above mechanisms are used for reducing the energy consumption of an individual worker. In the task allocation, several studies focus on how to optimize the overall energy consumption for MCS systems. For instance, to minimize the energy consumption, [34] attempts to minimize total energy consumption while ensuring the required spatial-temporal coverage. The authors in [37] formulate a task allocation problem, whose objective is to maximize sensing quality while minimizing energy consumption. The authors in [35] formulate another MCS task allocation problem, in which the objective is to maximize the task quality given a limited overall energy consumption. The study in [53] formulates another version of task allocation problem by considering the energy consumption, worker' reputation, and budget limitation all together.

### B. Single-Objective Allocation VS Multi-Objective Allocation

Most of the existing research works formulate the MCS task allocation as a single objective optimization problem, in which they only aim at optimizing one specific goal while keeping others as constraints. For example, the formulated problems in literature such as [34,35,43,44,45,46] are all single-objective-oriented. On the other hand, some others formulate the MCS task allocation as a multi-objective optimization problem [33,37]. For example, [37] aims at maximizing sensing quality while minimizing energy consumption in MCS task allocation. The objective of [33] is to minimize the traveling cost and meanwhile maximizing the number of completed MCS tasks. The multi-objective optimization problems formulated in [33,37] are commonly transformed into single-objective optimization problems based on the theory in [38], in which the weight of each objective is defined by task organizer. The shortcoming of such transformation is that sometimes it is difficult for the task organizers to decide the weight parameters.

### C. Single-Task-Oriented Allocation VS Multi-Task-Oriented Allocation

In the earlier stage of MCS research, existing approaches (e.g., [34,35,37,40,43,44]) are mostly single-task oriented, where they assume that tasks on MCS platforms are isolated so that the task allocation is executed for each single task independently. However, as the number of MCS tasks increases, the tasks are no longer independent, because they compete with each other in a shared and limited resource pool (e.g., shared user pool or total budget). Thus, in order to better coordinate tasks and make full use of the limited resources, some recent studies (e.g., [50,51,52,79]) have started to focus on multi-task allocation, where the interdependency of multiple tasks is considered. Typically, the objective is to optimize the overall utility of multiple tasks. For example, [52,79] studied the overall utility maximization of multiple tasks with worker's sensing capability constraints, while [33, 50, 51] proposed frameworks to optimize the overall utility with a total incentive budget constraint. In these works, the overall utility is all defined as the weighted sum of each task's sensing quality (e.g., spatial-temporal coverage).

### D. Offline Allocation VS Online Allocation

In terms of the timing when the allocation solution is determined, MCS tasks allocation can be either online or offline. If the tasks are assigned before the start time of the MCS task execution, it is the offline mode. On the contrary, if the task allocation is performed while the MCS task is running, it is the online allocation. For example, studies such as [34,35,43,51,52] are based on offline mode. The offline mode does not require the workers' real-time location information, which is more privacy-preserving. However, one main technical challenge for offline task allocation is that the system should be able to predict the workers' mobility accurately based on historical records. In contrast, existing studies, such as [62,63], adopt the online mode. The objective of [62] is to minimize the number of redundant task assignments while ensuring the required number of participants returning the sensing results within each time slot. A study in [63] aims at minimizing the number of assigned tasks while ensuring the full coverage the target area in each time slot. Compared to the offline mode, online task allocation has more knowledge about the real-time location of worker $u$ in time slot $i$, if $u$ uploads data with geotagging in previous time slots $(1,2…i-1)$. Thus, the mobility prediction can be easier with the combination of both real-time location and historical mobility records.

## V. MCS TASK ALLOCATION ALGORITHMS

### A. General Framework

Though with different goals and constraints, the task allocation can be formulated as combinatorial optimization problems, which attempt to find an optimal solution from a large search space. For instance, several studies aim to find a subset set of workers [34,35,47,50], while some others' goal is to find a subset of task-and-worker pairs [32,33,51,52,55,56]. Intuitively, it is easy to think of a brute-force approach, where it can estimate the utility of each possible combination so that the optimal one can be obtained. However, the formulated combinatorial optimization problems are usually NP-hard, thus the brute force approach is not acceptable when there are a large number of workers or tasks. Therefore, existing research work commonly chooses to design approximation allocation algorithms to achieve the near-optimal solution, which can be divided into the following two categories.

The general framework for MCS task allocation is presented in Fig. 4, which consists of two major components: (1) *Utility Estimation*: the algorithms for estimating the utility of a given set (a set of workers or task-and-worker pairs). Usually, the estimation needs the understanding of the workers' mobility pattern so that the historical mobility records profiling and mobility prediction are the basic components. (2) *Searching Process*: the searching algorithms to obtain a near-optimal solution. The algorithms are divided as greedy or non-greedy in this article.

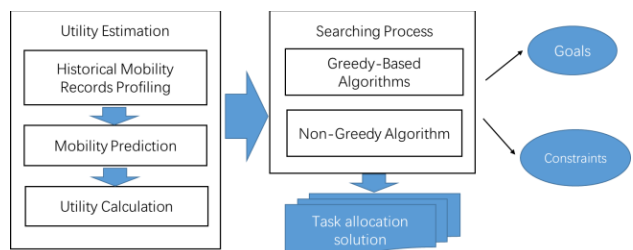

Fig. 4 General framework for MCS task allocation

## B. Greedy Based Algorithms

Most of the existing studies for MCS task allocation adopt the greedy based algorithms [34,35,43,44,50,51, 52], in which it iteratively selects "best" element (i.e., a worker or a task-and-worker pair) and adds into a set until certain stopping criterion is triggered (e.g., budget is used up, required coverage is reached, or none of the workers can be assigned to new tasks). After the greedy process stops, the obtained set is the near-optimal solution. For example, the greedy-based algorithm in [34] iteratively selects the participant with the maximum estimated coverage increase until the coverage requirement is satisfied. In some special problem settings such as [35,52], as the utility function cannot be estimated directly, multiple rounds of the greedy process should be executed to improve the optimality. In the experimental evaluation, the proposed greedy based approaches are proved to be effective in their targeted application scenarios under various settings.

However, in terms of theoretical bound, some algorithms have approximation bound guarantees, while others do not. Whether the proposed greedy algorithms have approximation bound are determined by the property of the defined utility functions and constraints. For example, the utility functions defined in [34,35,43,44,50,51] are the submodular set function with cardinality constraints, so the greedy based approaches can achieve at least 1-1/e ($\approx 0.63$) approximation bound compared to the optimal solution. We look at the exemplary problem, which is defined as selecting a fixed number of users with the objective of maximizing coverage. In this example, with a limited number of workers (e.g., 1000 workers), if the optimal solution (e.g., the brute-force approach) can get a coverage of 0.98, then the near-optimal solution can get at least a coverage of 0.9*0.63=0.617 even in the worst case. On the contrary, the utility functions used in [32,33,52] are non-submodular, so that the strict approximation bound is not declared in these studies. Considering that greedy based algorithms enjoy a good empirical performance in studies such as [52], in which the utility functions are non-submodular, it is interesting to investigate if a certain theoretical bound exists in future work. Recent literature regarding the guarantees for greedy optimization of non-submodular functions [54] may inspire us to address this issue.

## C. Non-Greedy Algorithms

Although through empirical studies the greedy based approaches are proved to be effective in their formulated problems and settings, they are not the skeleton key for all rooms as the formulation of MCS task allocation is diversifying. The greedy algorithms are sub-optimal in some scenario because they select the local best at each step. Therefore, more sophisticated algorithms have been designed.

For example, genetic algorithms (GA) are used in [32] for optimizing time-sensitive and time-tolerant MCS task allocation problems. In GA, through several generations of selection, crossover and mutation, the initial population (i.e., initial task allocation solution) converges to the optimal or near-optimal solution. The authors in [33] transform the problem using the Minimum Cost Maximum Flow (MCMF) theory and construct a new MCMF model by considering different constraints, then propose the MT-MCMF and MTP-MCMF algorithms. Both [55] and [56] formulate the MCS task allocation as a bipartite graph partition problem and propose divide-and-conquer algorithms. In order to maximize the number of tasks allocated to each worker, two algorithms are developed using dynamic programming and branch-and-bound strategies [57]. Also, a bisection-based algorithm[1] is developed in [58] that performs top-down recursive bisection and a bottom-up merge procedure iteratively so that assignment and scheduling can be performed locally in a much smaller promising space.

## D. Algorithm Evaluation

We should evaluate the task allocation algorithm before applying it in real-world MCS systems. The common strategy for evaluating the algorithms is to compare the performance with different baselines under various settings (e.g., the number of tasks and workers, workers' bandwidth, total incentive budget, task distribution, etc.). One of the biggest challenges for the MCS research community to evaluate the task allocation algorithm is the absence of public real-world datasets from applications. Therefore, the existing work always evaluates the algorithms' performance based on both the real-world and synthetic datasets. The information of workers' mobility is usually based on a real-world dataset (such as D4D [2] and Gowalla [72]), while the information of task (such as spatial-temporal distribution, budget, and required quality) are commonly synthetic. A typical example of the real-world dataset used is the D4D dataset [2], which contains two data types. One data type contains the information about cell towers, including tower id, latitude, and longitude. The other one contains 50,000 users' phone call records in Ivory Coast. The D4D dataset is used in the evaluation of task allocation algorithms such as [33,34,35,51,52]. For the synthetic dataset, one representative example is that the authors in [61] propose a toolbox to generate synthetic data for experimentation of MCS, thus leading to reproducible research.

Table 1 summarizes the characteristics of problem formulation and allocation algorithm for each MCS task allocation study. We hope this could help readers quickly identify the subset of relevant papers for his/her purposes.

## VI. FUTURE RESEARCH OPPORTUNITIES AND PROPOSALS

Existing work on MCS has studied various aspects for task allocation. However, the gap between ideal problem setting and real-world applications still prevent MCS system from being widely deployed. Thus, we next highlight several directions for future research by taking some practical issues into account.

---

[1] The bisection method is a root-finding method that repeatedly bisects an interval and then selects a subinterval in which a root must lie for further processing.

Table 1  A summary of characteristics of problem formulation and allocation algorithm for each analyzed paper

| reference | sensing quality metric | type of cost | single/ multi objective | single/multi task | online /offline | algorithm |
|---|---|---|---|---|---|---|
| 32 | spatial-temporal coverage based | incentive cost | single | multiple | online | Genetic Algorithms |
| 55 | spatial-temporal coverage based | incentive cost | single | multiple | online | divide-and-conquer |
| 57 | spatial-temporal coverage based | incentive cost | single | multiple | online | branch-and-bound |
| 33 | spatial-temporal coverage based | incentive cost | multiple | multiple | online | MCMF |
| 56 | spatial-temporal coverage based | incentive cost | multiple | multiple | online | divide-and-conquer |
| 58 | spatial-temporal coverage based | incentive cost | multiple | multiple | online | Bisection-based |
| 34,43,46 | spatial-temporal coverage based | incentive cost | single | single | offline | greedy |
| 35 | spatial-temporal coverage based | energy consumption | single | single | offline | greedy |
| 37,48 | spatial-temporal coverage based | energy consumption | multiple | single | offline | greedy |
| 40 | sensor data value based | incentive cost | multiple | multiple | online | matrix completion |
| 44 | sensor data value based | incentive cost | single | multiple | online | greedy |
| 45 | spatial-temporal coverage based | incentive cost | single | multiple | online | rule-based |
| 47 | spatial-temporal coverage based | incentive cost | multiple | single | offline | probabilistic registration |
| 49 | spatial-temporal coverage based | energy consumption | single | multiple | online | dynamic programming |
| 50,51,52 | spatial-temporal coverage based | incentive cost | single | single | offline | greedy |
| 53 | spatial-temporal coverage based | energy consumption | multiple | multiple | online | MMO |
| 62 | spatial-temporal coverage based | energy consumption | multiple | multiple | online | adaptive pace control |
| 63 | spatial-temporal coverage based | energy consumption | single | multiple | online | adaptive pace control |
| 66 | sensor data value based | location privacy | single | multiple | online | non-linear programming |
| 68 | sensor data value based | location privacy | multiple | multiple | online | probabilistic registration |

A. *Sustainable MCS Task Allocation.*

Existing studies usually focus on short-term task allocation in MCS. For instance, the organizer allocates the sensing task of traffic accident detection to participators immediately and the participators attempt to complete the tasks as soon as possible. In contrast, there are also many long-term sensing tasks, such as air quality surveillance for several years, which are significant for future cities and do need the sustainable task allocation. To achieve the sustainability, four directions are required to be considered. First, unlike the one-time budget in most of recent literature, a continuous investment/spending model should be formulated and its dynamic balance is valuable to be derived. Second, more attention should be payed to the participator experience. Not only the incentive mechanism but also the cultural recognition can motive the long-term participant. Third, 'Rome was not built in a day'. The penetration of MCS will be a gradual process. Current studies always assume that all users or a given probability of users would accept the task allocation, which is not applicable in practice. It is better to define a new feature of penetration to characterize the development of MCS.

Fourth, a green task allocation is valuable in sustainable MCS. Here, the 'green' have several meanings including: adopt the green energy, minimize the junk/redundant information, and reduce the human cost. Based on the above directions, we think sustainable MCS task allocation is still an uncovered treasure, worthy of our researching.

*B. Behavioral Models for Improving Task Allocation.*

Actually, many factors will affect users' behavior in task completion, which is crucial for task allocation. For example, if we can predict the workers' task acceptance likelihood, then we can further optimize the task allocation by assigning more tasks to those more likely to accept it [65]. Literatures of general crowdsourcing predict workers' behaviors by considering factors such as topical interest, expertise and time availability. In addition to that, MCS should further consider many other contextual factors. For example, contexts (e.g., the participants' motion and the position of the mobile device) has a significant impact on the sensing data quality for certain types of MCS tasks. We can train a sensing data quality classifier, which extract the relation between context information (such as the participants' motion) and sensing data quality, to estimate data quality in MCS. This classifier can be applied to guide user recruitment and task assignment in MCS. In another example, by detecting instances where a participant is bored, it is then possible to take advantage of their contextual cognitive surplus.

*C. Hybrid MCS Task Allocation*

Existing task allocation solutions adopt either the opportunistic mode or the participatory mode (mentioned in Section III). Motivated by the complementary nature of these two modes, there may be a hybrid solution, which can effectively integrate the opportunistic-mode and the participatory-mode task allocation. For example, we can recruit a number of opportunistic workers to complete tasks during their routine trajectories. Then, we further assign some other participatory workers to locations where tasks cannot be completed by the opportunistic workers alone. The hybrid solution has two advantages. First, from the perspective of the workers, it naturally accommodates the workers' participation preferences and makes full use of the available human sensing resources. Although the workers all want to contribute sensing data to MCS tasks, their preferred way of participation can be different. For example, some office employees are busy all day and do not have time to take a detour for task completion. In this case, they only accept to complete tasks on their daily routine trajectories. In contrast, some retired or unemployed citizens who have plenty of leisure time may be willing to move intentionally and complete tasks to earn incentive rewards. Second, from the perspective of the task organizer, it can achieve a better tradeoff between sensing quality and cost. Compared with pure participatory-mode approaches, it leverages some opportunistic workers to unintentionally complete tasks, which significantly reduces the incentive cost. In contrast to the pure opportunistic-mode approaches, it further improves the sensing quality by assigning some participatory workers to move and complete tasks in uncovered locations. However, when the task allocation of these two types of workers is correlated (e.g., they share a total incentive budget), it is challenging to jointly optimize them, which remains as a future research issue.

*D. Considering Data Sharing Among Multiple Tasks.*

Existing work for MCS task allocation only considers the competitive relation among multiple tasks. That is, if a sensing resource (workers) is allocated to some tasks, other tasks cannot utilize it. However, we can take into account more complicated situations, where sensing results for a task can be utilized for another task. Intuitively, although the number of sensing tasks may become larger and larger with the popularity of MCS, the kinds of sensors in the smartphone are limited. To this end, some tasks can share the same type of sensing data, or the sensing data among tasks are co-related. For example, the queue time estimation task in [69] needs to use GPS, accelerometers, and microphones, while noise level monitoring task requires GPS and microphones. In this case, the GPS and accelerometers can be shared.

*E. Social-Network-Assisted MCS Task Allocation*

Existing studies commonly recruit workers and allocate tasks on specialized MCS systems with assumed large user pools, so that their goal is to select a subset of users from the pool with the consideration of some factors (e.g., sensing quality and cost). However, they fail to work when such assumed large user pools do not exist. In the recent decade, the popularity of mobile social networks (MSN, e.g., Facebook, Twitter, and Foursquare, etc.) has created new mediums for information sharing and propagation, and they have gradually become promising platforms for advertising novel products or innovative ideas. Inspired by the power of MSN, instead of relying on specific MCS platforms, it is interesting to study how to recruit workers of MCS task in a novel manner, i.e., exploiting social network as the task allocation platform. Nevertheless, we cannot directly adopt the information propagation model of the social network in social-network assisted MCS task allocation. When determining whether the user will be influenced by the propagated information, the existing models merely consider the influence from the neighbors in the social network without taking the specific factors about MCS tasks into account. For example, whether the incentive is attractive or whether the task's topic is interesting would have a significant impact on the users' decision on accepting or declining the task. Thus, it needs to extend the state-of-the-art propagation models in the social network research community by introducing MCS-specific factors.

*F. Composite MCS Task Allocation*

For previous work of MCS task allocation, sensing tasks are rather simple, where a participant's mobile device can provide a complete sample by utilizing a single type of sensor. In the real-world application scenarios, however, there are some other MCS tasks which can be rather complex, which consists of several subtasks and different types of sensors or sensing capability. We refer such complicated tasks as the *composite*

MCS tasks. Air quality monitoring task is a typical example of composite MCS because the AQI (Air Quality Index) is calculated based on the sensor readings of multiple types of pollutants, including ground-level ozone, particulates, sulfur dioxide, carbon monoxide and nitrogen dioxide. A participant usually fails to provide a full sample for a composite task, because he/she may not have the sensing capabilities of all subtasks. For example, their smartphone may not be embedded with the required sensors (e.g., $SO_2$ sensor), or they deliberately disable the sensors (e.g., microphone) to preserve their privacy. If we assume that the mobile device of each participant is embedded with a subset of the required pollutant sensors, then a complete AQI in a certain place should be obtained through the collaborative sensing among multiple participants. As each participant is only able to complete a subset of sub-tasks, the composite task should be accomplished through the collaboration of multiple participants. Therefore, the task allocation of the composite task is much more complicated, so that the study on the task allocation of the composite MCS is an important direction for future research.

G. *Location-Privacy-Concerned MCS Task Allocation*

While much theoretical privacy protection has been proposed in MCS task allocation, it seems that in real applications, privacy protection is still often ignored, or implemented by some simple configuration options where users can set private locations to avoid being sensed. This phenomenon may be because users are often hard to understand the real privacy protection effect for them if the privacy mechanism is not intuitively comprehensible. Moreover, in reality, many users may be unclear about the potential consequences incurred by privacy leakage [64], which makes implementing privacy mechanisms is not urgent for MCS business entities. Therefore, there is still a huge gap between the industry and academia in the MCS location privacy concerned task allocation. To fill this gap, one possible direction is to design more user-friendly (understandable) privacy mechanisms and educate the public about the severe privacy leakage consequences, so as to make the users more concerned about their privacy and get the most appropriate privacy configurations; and another is to make some guidelines and regularizations about user privacy for MCS business entities, so as to facilitate a more secure sensing environment for MCS participants.

H. *Task Allocation for Sparse MCS*

While many MCS task allocation methods are proposed to maximize the sensing coverage of the target area, how to deal with the missing data of un-sensed regions are often neglected in those methods. Recently, researchers have proposed 'sparse MCS' paradigm, where the treatment of such missing data in un-sensed regions is formalized as an important stage. State-of-the-art machine learning approaches like matrix completion and compressive sensing are used in this stage to infer the missing data with high quality [40, 41]. In sparse MCS, the target of task allocation differs from coverage maximization, as the sensing data of different regions at different time slots can contribute diversely to the overall missing data inference quality. However, because the ground truth sensing values of un-sensed areas are unknown, how to quantify the data inference quality is really challenging. Rather than directly comparing the inferred data with ground truth, novel methods have to be developed to measure the data inference quality. If more real-life factors are added, e.g., different participants are paid with different incentives, the task allocation for sparse MCS will become even more complicated. To this end, how to design effective and efficient task allocation schemes for spare MCS needs more research efforts.

I. *Task Allocation for Indoor MCS.*

Existing task allocation approaches are mainly designed for outdoor scenarios. MCS in indoor areas is becoming more and more crucial for flow management, security and surveillance, or building usage statistics in recent years. For example, studies such as [76,77] proposed floor plan reconstruction and indoor navigation systems by leveraging crowd-sensed data from mobile users. These studies mainly focus on the inference of the floor layout or people's locations given a fixed set of mobile devices and their signals. It is interesting to further study the task allocation problem for these indoor MCS applications. For example, if the candidate users who are willing to share the signals require certain incentive reward, then it is interesting to study how to select a set of devices for jointly optimizing the incentive cost and accuracy of floor reconstruction or indoor navigation.

VII. CONCLUSION

In this article, we survey the task allocation problem of a special case of crowdsourcing, named mobile crowd sensing, which requires workers' physical presence at certain locations in order to complete urban environment sensing tasks. We discuss the unique characteristics of MCS. We then classify the state-of-the-art research into different categories with different problem formulation or allocation algorithms. In the end, we suggest several promising issues as future research directions.

REFERENCES

1. Zanella, A., Bui, N., Castellani, A., Vangelista, L., & Zorzi, M. (2014). Internet of things for smart cities. IEEE Internet of Things journal, 1(1), 22-32.

2. Antonić, A., Marjanović, M., Pripužić, K., & Žarko, I. P. (2016). A mobile crowd sensing ecosystem enabled by CUPUS: Cloud-based publish/subscribe middleware for the Internet of Things. Future Generation Computer Systems, 56, 607-622.

3. An, J., Gui, X., Wang, Z., Yang, J., & He, X. (2015). A crowdsourcing assignment model based on mobile crowd sensing in the internet of things. IEEE Internet of Things Journal, 2(5), 358-369.

4. R.K. Ganti, F. Ye, and H. Lei. Mobile crowdsensing: Current state and future challenges. IEEE Communications Magazine, 49:32–39, 2011.

5. Burke, J. A., Estrin, D., Hansen, M., Parker, A., Ramanathan, N., Reddy, S., & Srivastava, M. B. (2006). Participatory sensing.

6. Alt, Florian, et al. "Location-based crowdsourcing: extending crowdsourcing to the real world." Proceedings of the 6th Nordic Conference on Human-Computer Interaction: Extending Boundaries. ACM, 2010.
7. Feng, Zhenni, et al. "TRAC: Truthful auction for location-aware collaborative sensing in mobile crowdsourcing." INFOCOM, 2014 Proceedings IEEE. IEEE, 2014.
8. Rogstadius, Jakob, et al. "CrisisTracker: Crowdsourced social media curation for disaster awareness." IBM Journal of Research and Development 57.5 (2013): 4-1
9. M. Ra, B. Liu, T. F. L. Porta, and R. Govindan. 2012. Medusa: A programming framework for crowd-sensing applications. In Proceedings of the 10th international conference on Mobile systems, applications, and services (MobiSys '12), 337-350
10. T. Das, P. Mohan, V. N. Padmanabhan, R. Ramjee, and A. Sharma. 2010. PRISM: platform for remote sensing using smartphones. In Proceedings of the 8th international conference on Mobile systems, applications, and services (MobiSys '10), 63-76.
11. Howe, J. (2006). The rise of crowdsourcing. Wired magazine, 14(6), 1-4.
12. Mao, K., Capra, L., Harman, M., & Jia, Y. (2017). A survey of the use of crowdsourcing in software engineering. Journal of Systems and Software,126, 57-84.
13. Xintong, G., Hongzhi, W., Song, Y., & Hong, G. (2014). Brief survey of crowdsourcing for data mining. Expert Systems with Applications, 41(17), 7987-7994.
14. Paulheim, H. (2017). Knowledge graph refinement: A survey of approaches and evaluation methods. Semantic web, 8(3), 489-508.
15. Kulkarni, C., Dow, S. P., & Klemmer, S. R. (2014). Early and repeated exposure to examples improves creative work. In Design thinking research(pp. 49-62). Springer International Publishing.
16. Mollick, E. (2014). The dynamics of crowdfunding: An exploratory study.Journal of business venturing, 29(1), 1-16.
17. Burke, J. A., Estrin, D., Hansen, M., Parker, A., Ramanathan, N., Reddy, S., & Srivastava, M. B. (2006). Participatory sensing. Center for Embedded Network Sensing.
18. Campbell, A. T., Eisenman, S. B., Lane, N. D., Miluzzo, E., Peterson, R. A., Lu, H., ... & Ahn, G. S. (2008). The rise of people-centric sensing. IEEE Internet Computing, 12(4).
19. Rana, R. K., Chou, C. T., Kanhere, S. S., Bulusu, N., & Hu, W. (2010, April). Ear-phone: an end-to-end participatory urban noise mapping system. In Proceedings of the 9th ACM/IEEE International Conference on Information Processing in Sensor Networks (pp. 105-116). ACM.
20. Mun, M., Reddy, S., Shilton, K., Yau, N., Burke, J., Estrin, D., ... & Boda, P. (2009, June). PEIR, the personal environmental impact report, as a platform for participatory sensing systems research. In Proceedings of the 7th international conference on Mobile systems, applications, and services (pp. 55-68). ACM.
21. Hull, B., Bychkovsky, V., Zhang, Y., Chen, K., Goraczko, M., Miu, A., ... & Madden, S. (2006, October). CarTel: a distributed mobile sensor computing system. In Proceedings of the 4th international conference on Embedded networked sensor systems (pp. 125-138). ACM.
22. Chon, Y., Lane, N. D., Kim, Y., Zhao, F., & Cha, H. (2013, September). Understanding the coverage and scalability of place-centric crowdsensing. In Proceedings of the 2013 ACM international joint conference on Pervasive and ubiquitous computing (pp. 3-12). ACM.
23. K. K. Rachuri, C. Mascolo, M. Musolesi, and P. J. Rentfrow, "SociableSense: exploring the trade-offs of adaptive sampling and computation offloading for social sensing," in MobiCom, 2011, pp. 73–84.
24. Reddy, S., Parker, A., Hyman, J., Burke, J., Estrin, D., & Hansen, M. (2007, June). Image browsing, processing, and clustering for participatory sensing: lessons from a DietSense prototype. In Proceedings of the 4th workshop on Embedded networked sensors (pp. 13-17). ACM.
25. Zhang, D., Wang, L., Xiong, H., & Guo, B. (2014). 4W1H in mobile crowd sensing. IEEE Communications Magazine, 52(8), 42-48.
26. Wang, J., Wang, Y., Helal, S., & Zhang, D. (2016). A Context-Driven Worker Selection Framework for Crowd-Sensing. International Journal of Distributed Sensor Networks, 12(3), 6958710.
27. Wang J, Wang Y, Zhang D, et al. Energy Saving Techniques in Mobile Crowd Sensing: Current State and Future Opportunities. IEEE Communications Magazine, 2018, 56(5): 164-169.
28. Lane, N. D., Chon, Y., Zhou, L., Zhang, Y., Li, F., Kim, D., ... & Cha, H. (2013, November). Piggyback CrowdSensing (PCS): energy efficient crowdsourcing of mobile sensor data by exploiting smartphone app opportunities. In Proceedings of the 11th ACM Conference on Embedded Networked Sensor Systems (p. 7). ACM.
29. Wang, Y., Lin, J., Annavaram, M., Jacobson, Q. A., Hong, J., Krishnamachari, B., & Sadeh, N. (2009, June). A framework of energy efficient mobile sensing for automatic user state recognition. In Proceedings of the 7th international conference on Mobile systems, applications, and services (pp. 179-192). ACM.
30. Guo, B., Wang, Z., Yu, Z., Wang, Y., Yen, N. Y., Huang, R., & Zhou, X. (2015). Mobile crowd sensing and computing: The review of an emerging human-powered sensing paradigm. ACM Computing Surveys (CSUR), 48(1), 7.
31. Yuen, M. C., King, I., & Leung, K. S. (2011, October). A survey of crowdsourcing systems. In Privacy, Security, Risk and Trust (PASSAT) and 2011 IEEE Third Inernational Conference on Social Computing (SocialCom), 2011 IEEE Third International Conference on (pp. 766-773). IEEE.
32. Guo, B., Liu, Y., Wu, W., Yu, Z., & Han, Q. (2016). ActiveCrowd: A Framework for Optimized Multitask Allocation in Mobile Crowdsensing Systems. IEEE Transactions on Human-Machine Systems.
33. Liu, Y., Guo, B., Wang, Y., Wu, W., Yu, Z., & Zhang, D. (2016, September). TaskMe: multi-task allocation in mobile crowd sensing. In Proceedings of the 2016 ACM International Joint Conference on Pervasive and Ubiquitous Computing (pp. 403-414). ACM.
34. Zhang, D., Xiong, H., Wang, L., & Chen, G. (2014, September). CrowdRecruiter: selecting participants for piggyback crowdsensing under probabilistic coverage constraint. In Proceedings of the 2014 ACM International Joint Conference on Pervasive and Ubiquitous Computing (pp. 703-714). ACM.

35. Xiong, H., Zhang, D., Chen, G., Wang, L., Gauthier, V., & Barnes, L. E. (2016). iCrowd: Near-optimal task allocation for piggyback crowdsensing. IEEE Transactions on Mobile Computing, 15(8), 2010-2022.
36. Wazir Zada Khan, Y. X., Aalsalem, M. Y., & Arshad, Q. (2013). Mobile phone sensing systems: A survey. IEEE Communications Surveys Tutorials, 15(1), 402-427.
37. Liu, C. H., Zhang, B., Su, X., Ma, J., Wang, W., & Leung, K. K. (2015). Energy-aware participant selection for smartphone-enabled mobile crowd sensing. IEEE Systems Journal.
38. R. T. Marler and J. S. Arora, "The weighted sum method for multi-objective optimization: new insights," Structural and multidisciplinary optimization, vol. 41, no. 6, pp. 853–862, 2010.
39. Y. Zhu, Z. Li, H. Zhu, M. Li, and Q. Zhang, "A compressive sensing approach to urban traffic estimation with probe vehicles," IEEE Transactions on Mobile Computing, vol. 12, no. 11, pp. 2289–2302, 2013
40. Wang, L., Zhang, D., Wang, Y., Chen, C., Han, X., & M'hamed, A. (2016). Sparse mobile crowdsensing: challenges and opportunities. IEEE Communications Magazine, 54(7), 161-167.
41. Wang, L., Zhang, D., Pathak, A., Chen, C., Xiong, H., Yang, D., & Wang, Y. (2015, September). CCS-TA: quality-guaranteed online task allocation in compressive crowdsensing. In Proceedings of the 2015 ACM International Joint Conference on Pervasive and Ubiquitous Computing (pp. 683-694). ACM.
42. Xu, Q., & Zheng, R. When Data Acquisition Meets Data Analytics: A Distributed Active Learning Framework for Optimal Budgeted Mobile Crowdsensing. INFOCOM 2017.
43. S. Reddy, D. Estrin, and M. Srivastava. Recruitment framework for participatory sensing data collections. In Proceedings of Pervasive, pages 138–155. 2010.
44. Adish Singla and Andreas Krause. Incentives for privacy tradeoff in community sensing. In First AAAI Conference on Human Computation and Crowdsourcing, 2013.
45. Giuseppe Cardone, Luca Foschini, Paolo Bellavista, Antonio Corradi, Cristian Borcea, Manoop Talasila, and Reza Curtmola. Fostering participaction in smart cities: a geo-social crowdsensing platform. Communications Magazine, IEEE, 51(6), 2013.
46. Karaliopoulos, M., Telelis, O., & Koutsopoulos, I. (2015). User Recruitment for Mobile Crowdsensing over Opportunistic Networks. In INFOCOM 2015 Proceedings, IEEE
47. S. Hachem, A. Pathak, and V. Issarny. Probabilistic registration for large-scale mobile participatory sensing. In Proceedings of the 2013 IEEE International conference on Pervasive Computing and Communications, volume 18, page 22, 2013.
48. Wang, L., Zhang, D., Yan, Z., Xiong, H., & Xie, B. (2015). effSense: a novel mobile crowd-sensing framework for energy-efficient and cost-effective data uploading. IEEE Transactions on Systems, Man, and Cybernetics: Systems, 45(12), 1549-1563.
49. Lane, N. D., Chon, Y., Zhou, L., Zhang, Y., Li, F., Kim, D., ... & Cha, H. (2013, November). Piggyback CrowdSensing (PCS): energy efficient crowdsourcing of mobile sensor data by exploiting smartphone app opportunities. In Proceedings of the 11th ACM Conference on Embedded Networked Sensor Systems (p. 7). ACM.
50. Z. Song, C. H. Liu, J. Wu, J. Ma, and W. Wang. 2014. QoI-Aware Multitask-Oriented Dynamic Participant Selection with Budget Constraints. IEEE Transactions on Vehicular Technology, 63: 4618-4632.
51. Wang, J., Wang, Y., Zhang, D., Xiong, H., Wang, L., & Sumi, H., et al. (2016). Fine-grained multi-task allocation for participatory sensing with a shared budget. Internet of Things Journal (in press).
52. Wang, J., Wang, Y., Zhang, D., Wang, F., He, Y., & Ma, L. PSAllocator: Multi-Task Allocation for Participatory Sensing with Sensing Capability Constraints. The, ACM Conference on Computer- Supported Cooperative Work and Social Computing (CSCW 2017).
53. Wang, W., Gao, H., Liu, C. H., & Leung, K. K. (2016). Credible and energy-aware participant selection with limited task budget for mobile crowd sensing. Ad Hoc Networks, 43, 56-70.
54. Bian, A. A., Buhmann, J. M., Krause, A., & Tschiatschek, S. (2017). Guarantees for Greedy Maximization of Non-submodular Functions with Applications. arXiv preprint arXiv:1703.02100.
55. Cheng, P., Lian, X., Chen, L., Han, J., & Zhao, J. (2016). Task assignment on multi-skill oriented spatial crowdsourcing. IEEE Transactions on Knowledge and Data Engineering, 28(8), 2201-2215.
56. Cheng, P., Lian, X., Chen, Z., Fu, R., Chen, L., Han, J., & Zhao, J. (2015). Reliable diversity-based spatial crowdsourcing by moving workers. Proceedings of the VLDB Endowment, 8(10), 1022-1033.
57. D. Deng, C. Shahabi, and U. Demiryurek, "Maximizing the Number of Worker's Self-Selected Tasks in Spatial Crowdsourcing," Proc. 21st ACM SIGSPATIAL Int'l. Conf. Advances in Geographic Information Systems, 2013, pp. 324–33.
58. D. Deng, C. Shahabi, and L. Zhu, "Task Matching and Scheduling for Multiple Workers in Spatial Crowdsourcing," Proc. 23rd Int'l. Conf. Advances in Geographic Information Systems, 2015, pp. 21:1--21:10.
59. Ibrahim, M., Chamoun, M., Kilany, R., El Helou, M., & Rouhana, N. (2017). Comiqual: collaborative measurement of internet quality. *Annals of Telecommunications*, 1-13.
60. V.D. Blondel, M. Esch, C. Chan, F. Clerot, P. Deville, E. Huens, F. Morlot, Z. Smoreda, and C. Ziemlicki. 2012. Data for development: the d4d challenge on mobile phone data. arXiv preprint arXiv:1210.0137.
61. To, H., Asghari, M., Deng, D., & Shahabi, C. (2016, March). SCAWG: A toolbox for generating synthetic workload for spatial crowdsourcing. InPervasive Computing and Communication Workshops (PerCom Workshops), 2016 IEEE International Conference on (pp. 1-6). IEEE.
62. Xiong, H., Zhang, D., Wang, L., Gibson, J. P., & Zhu, J. (2015). EEMC: Enabling energy-efficient mobile crowdsensing with anonymous participants. ACM Transactions on Intelligent Systems and Technology (TIST), 6(3), 39.
63. Xiong, H., Zhang, D., Wang, L., & Chaouchi, H. (2015). Emc$^3$: Energy-efficient data transfer in mobile crowdsensing under full coverage constraint. IEEE Transactions on Mobile Computing, 14(7), 1355-1368.
64. Acquisti, A., Brandimarte, L., & Loewenstein, G. (2015). Privacy and human behavior in the age of information. Science, 347(6221), 509-514.

65. Karaliopoulos, M., Koutsopoulos, I., & Titsias, M. (2016, July). First learn then earn: Optimizing mobile crowdsensing campaigns through data-driven user profiling. In Proceedings of the 17th ACM International Symposium on Mobile Ad Hoc Networking and Computing (pp. 271-280). ACM.
66. Wang, L., Yang, D., Han, X., Wang, T., Zhang, D., & Ma, X. (2017). Location Privacy-Preserving Task Allocation for Mobile Crowdsensing with Differential Geo-Obfuscation. Proc. of the 26th International Conference on World Wide Web, pp. 627-636.
67. Pournajaf, L., Garcia-Ulloa, D. A., Xiong, L., & Sunderam, V. (2016). Participant privacy in mobile crowd sensing task management: a survey of methods and challenges. ACM SIGMOD Record, 44(4), 23-34.
68. Vergara-Laurens, I. J., Mendez, D., & Labrador, M. A. (2014). Privacy, quality of information, and energy consumption in participatory sensing systems. In Pervasive Computing and Communications (PerCom), 2014 IEEE International Conference on, pp. 199-207.
69. Wang, J., Wang, Y., Zhang, D., Wang, L., Chen, C., Lee, J. W., & He, Y. (2016). Real-time and generic queue time estimation based on mobile crowdsensing. Frontiers of Computer Science, 2017.11(1), 49-60
70. M. Duckham and L. Kulik, "A formal model of obfuscation and negotiation for location privacy," in Pervasive Computing. Springer, 2005, pp. 152–170.
71. Wang, L., Zhang, D., Yang, D., Lim, B.Y. and Ma, X., 2016, December. Differential location privacy for sparse mobile crowdsensing. In Data Mining (ICDM), 2016 IEEE 16th International Conference on (pp. 1257-1262). IEEE.
72. Scellato, S., Noulas, A., & Mascolo, C. (2011, August). Exploiting place features in link prediction on location-based social networks. In Proceedings of the 17th ACM SIGKDD international conference on Knowledge discovery and data mining (pp. 1046-1054). ACM.
73. Cheng, P., Lian, X., Chen, L., & Shahabi, C. (2017, April). Prediction-Based Task Assignment in Spatial Crowdsourcing. In Data Engineering (ICDE), 2017 IEEE 33rd International Conference on (pp. 997-1008). IEEE.
74. To, H., Ghinita, G., Fan, L., & Shahabi, C. (2017). Differentially private location protection for worker datasets in spatial crowdsourcing. IEEE Transactions on Mobile Computing, 16(4), 934-949.
75. Leye Wang, Daqing Zhang, Dingqi Yang, Animesh Pathak, Chao Chen, Xiao Han, Haoyi Xiong, Yasha Wang (2017). SPACE-TA: Cost-Effective Task Allocation Exploiting Intradata and Interdata Correlations in Sparse Crowdsensing. ACM Transactions on Intelligent Systems and Technology, accepted.
76. Gao, R., Zhao, M., Ye, T., Ye, F., Wang, Y., Bian, K., ... & Li, X. (2014, September). Jigsaw: Indoor floor plan reconstruction via mobile crowdsensing. In Proceedings of the 20th annual international conference on Mobile computing and networking (pp. 249-260). ACM.
77. Zhang, C., Subbu, K. P., Luo, J., & Wu, J. (2015). GROPING: Geomagnetism and crowdsensing powered indoor navigation. IEEE Transactions on Mobile Computing, 14(2), 387-400.
78. Kim K, Zabihi H, Kim H, et al. TrailSense: A Crowdsensing System for Detecting Risky Mountain Trail Segments with Walking Pattern Analysis. Proceedings of the ACM on Interactive, Mobile, Wearable and Ubiquitous Technologies, 2017, 1(3): 65.
79. Wang J, Wang Y, Zhang D, et al. Multi-Task Allocation in Mobile Crowd Sensing with Individual Task Quality Assurance. IEEE Transactions on Mobile Computing, 2018.

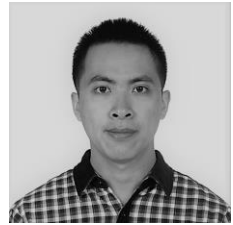

**Jiangtao Wang** received his Ph.D. degree in Peking University, Beijing, China, in 2015. He is currently an assistant professor in Institute of Software, School of Electronics Engineering and Computer Science, Peking University. His research interest includes mobile crowd sensing, collaborative computing, and social computing.

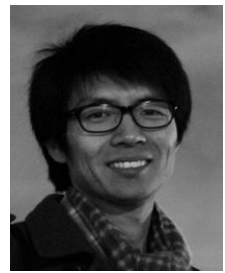

**Leye Wang** is currently a postdoctoral research fellow in Hong Kong University of Science and Technology. He obtained his Ph.D. from Institut Mines-Télécom/Télécom SudParis and Université Pierre et Marie Curie, France, in 2016. He received his M.Sc. and B.Sc. in computer science from Peking University, China. His research interests include mobile crowdsensing, social networks, and intelligent transportation systems.

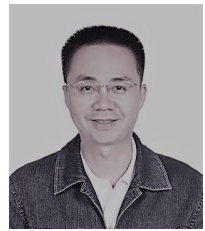

**Yasha Wang** received his Ph.D. degree in Northeastern University, Shenyang, China, in 2003. He is a professor and associate director of National Research & Engineering Center of Software Engineering in Peking University, China. His research interest includes urban data analytics, ubiquitous computing, software reuse, and online software development environment.

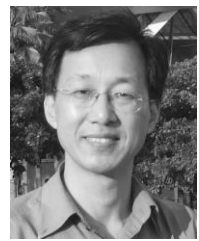

**Daqing Zhang** is a professor at Peking University, China, and Télécom SudParis, France. He obtained his Ph.D from the University of Rome "La Sapienza," Italy, in 1996. His research interests include context-aware computing, urban computing, mobile computing, and so on.

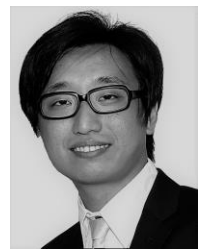

**Linghe KONG** is currently a tenure-track research professor with Department of Computer Science and Engineering at Shanghai Jiao Tong University. He received his Ph.D. degree in computer science from Shanghai Jiao Tong University, China, 2012. His research interests include wireless networks, 5G communication, big data, mobile computing, Internet of things, and smart energy systems.